# Entanglement of Quantum States, Thermodynamical Statistical Distributions and Physical Nature of Temperature

A.Yu. Bogdanov, Yu.I. Bogdanov, K.A. Valiev
(e-mail: bogdan@ftian.oivta.ru)

*Institute of Physics and Technology, Russian Academy of Sciences*

Thermodynamical equilibrium is considered as an effect of quantum entangling of the vacuum state of the system. An explicit mathematical model of multi-particle entangled pure quantum states is developed and analyzed. In the framework, the process of measurement results in probability distributions that exactly correspond to the heat equilibrium of a system in a thermostat.

**Introduction**

> "First of all, we need to define one of the principal notions of the general theory of heat – the notion of temperature - which is still far from being properly described" [1],
> L. Boltzmann "To the mechanical meaning of the second principle of heat theory" - 1866.
> (Wien. Ber., 1866, Bd.53, S.195-220)

The present study is closely related to our recent work [2], where the notion of information was introduced based on Schmidt decomposition. It was shown that the introduced notion of information described the measure of non-randomness of correlations between two observers that conduct measurements upon EPR (Einstein-Podolsky-Rosen) quantum states. By example of a two-dimensional normal distribution that corresponded to the vacuum state of two coupled oscillators, it was demonstrated that the initial pure quantum state, characterized by the entanglement of the two systems, after the measurement yielded probability distribution that corresponded to the heat equilibrium. In statistical physics, however, multi-particle quantum states are of more interest than two-particle states. The present work is devoted to that problem. Below, an explicit multi-particle quantum model will be analyzed. In the model entanglement takes place as oscillator particles interact by a mutual intermediary (partition or wall). The proposed model allows one to study the nature of equilibrium quantum distributions. It is shown that in the described system heat equilibriums arise during the process of measurement. The temperature of the system is defined as the ratio of the mass of the particles to the mass of the partition.

    The question of the nature of statistical distributions has been widely discussed since Boltzmann [1]. In the problem of mechanical description of heat theory the views of Boltzmann have passed two periods. (see the editor's article in [1]) . At first, the main issue was to combine mechanics with atomic science. Then, statistical aspects of the theory became of most importance. The famous H-theorem that Boltzmann formulated in 1872 was argued by many critics. Here, the principal problem is that due to the time reversibility of the laws of mechanics, a symbiosis of classical mechanics and statistics inevitably becomes internally contradictory and inconsequential. For instance, Poincare criticizes the grounds of Maxwell-Boltzmann statistical thermodynamics and claims that: "there is no need to argue to doubt a reasoning where the premise contradicts with the conclusion in respect to reversibility" (see Poincare «Mechanical view and experience» (1893) in [1]). Other arguments against Boltzmann's statistical mechanics were by Loschmidt (Boltzmann's teacher and colleague) and Zermelo (Planck's assistant). The arguments were also based on the reversibility of the laws of classical mechanics (see [1]). Boltzmann's answer was that H-theorem was statistical and it only described the most probable

path of system evolution, but did not exclude less probable events (fluctuations). Though, the answer was quite convincing from the practical view it still raised doubts methodologically.

It is worth noting that according to the modern scientific approach Boltzmann's statistical thermodynamics was not pure classical. Even in his famous work in 1872 "Further research of heat equilibrium between gas molecules" where he formulated the H-theorem, Boltzmann used quantum notions of energy (see [1] for more details). Still, he treated the quantum notion of energy only as an artificial mathematical tool and so did Planck 28 years later. Only after the works by Bohr, Heisenberg and Schrödinger did the quantum nature of energy became treated as a real physical property.

Therefore, there has never been pure classical statistical mechanics as a completed field of physics. All approaches to combine mechanics and statistics led to quantum representations (though often implicitly as for Boltzmann and Planck).

In quantum mechanics the problem of the reversibility is much different from classical mechanics [3]. Despite the fact that the Schrödinger equation is symmetric in time it includes some non-equivalence in time directions. Such non-equivalence in time directions is due to the fact that quantum state is reduced during the measurement. Quantum mechanics can be considered as a generalization of classical mechanics where statistical laws are fundamental [4].

Entanglement is one of the principal notions of quantum informatics. It is an important source for quantum computations and quantum cryptography. The role of entanglement in the problem of quantum state measurement has been studied by many authors. (see detailed reviews in [5],[6],[7]). Accounting for the entanglement between degrees of freedom of quantum bits (qubits) plays a vital role in the problem of enhancing quality of quantum informational systems [8,9].

It seems that the considered matters are of much importance both fundamentally and practically. Fundamentally, it is important that quantum mechanics is a statistical theory. In that sense quantum theory includes the grounds for statistical physics development. In other words, there are good prerequisites for a symbiosis of quantum mechanics and statistics. In contrast to the symbiosis between classical physics and statistics that is internally contradictory.

In principle, the models of quantum statistical physics have to be the models of quantum mechanics where the number of particles does not have to be small. Thus, according to [3], it is natural to suppose that quantum mechanics includes the principal basis for statistical physics description. In the present work we consider a simple (but multi-particle) model where the state of heat equilibrium is the result of measurement of an entangled multi-particle state.

It appears that better understanding of the notion of entanglement is important for applications. According to quantum informatics the issue studied in the work is the process of "heating" of a system during the measurement. Such a process (if it is uncontrolled) leads to the decoherence of the system state and to the loss of quality.

The work is divided into two parts. In the first part, multi-particle entangled EPR-type states are constructed and their statistical properties are studied. In the second part, a physical model based on oscillators is presented, where such EPR-states can arise.

1. **Multi-particle EPR-type states**

The principal goal of this section is to construct $r+1$- particle states of the following form:

$$\psi(x_1,..,x_{r+1}) = \sum_{n_1,...,n_r=0}^{\infty} c_{n_1,...,n_r} \psi_{n_1}^{(1)}(x_1)...\psi_{n_r}^{(r)}(x_r)\psi_{n_1+...+n_r}^{(r+1)}(x_{r+1}) \quad (1)$$

These are EPR-type states. Here one of the particles ($r+1$- th) is distinguished.

The results of measurements upon $r$ particles allow one to predict the state of the $r+1$-th particle (Its' number is equal to the sum of numbers of other particles). If during the process some other particle (not the $r+1$-th) is left unmeasured then one has to subtract from the $r+1$- th state number the sum of all other particles' state numbers. In a particular case when we measure a distinguished particle ($r+1$-th) and it appears to be in the ground state then all other particles are also bound to be in the ground state.

Let us introduce $r$ real parameters $f_1, f_2, ..., f_r$. Then to every point inside the hyper-sphere of a unity radius $f_1^2 + f_2^2 + ... + f_r^2 < 1$ we may put into correspondence some quantum state as described below.

Let $f^2 = f_1^2 + f_2^2 + ... + f_r^2$ - be the square of the parameter vector length

Then the decomposition coefficients and their squares (weights) are

$$c_{n_1,n_2...n_r} = (f_1)^{n_1}(f_2)^{n_2}...(f_r)^{n_r}\sqrt{(1-f^2)\frac{(n_1+n_2+...+n_r)!}{n_1!n_2!...n_r!}} \quad (2)$$

$$\lambda_{n_1,n_2...n_r} = (c_{n_1,n_2...n_r})^2 = (1-f^2)(f_1^2)^{n_1}(f_2^2)^{n_2}...(f_r^2)^{n_r}\frac{(n_1+n_2+...+n_r)!}{n_1!n_2!...n_r!} \quad (3)$$

The introduced weight coefficient satisfy the normalization condition

$$\sum_{n_1,n_2,...,n_r=0}^{\infty} \lambda_{n_1,n_2,...,n_r} = 1 \quad (4)$$

For the sake of simplicity of notions let us assume that the base functions of all $r+1$ particles have the following most simple form of oscillator base functions

$$\psi_k^{(j)}(x_j) = \frac{1}{(2^k k!\sqrt{\pi})^{1/2}} H_k(x_j)\exp\left(-\frac{x_j^2}{2}\right), \quad j=1,2,..,r+1, \quad k=0,1,2,... \quad (5)$$

By introducing scale multipliers we may obviously generalize the base functions.
Let us use the equation for Hermit-polynomials sum [10]:

$$\frac{\left(\sum_{k=1}^r f_k^2\right)^{n/2}}{n!} H_n\left(\frac{\sum_{k=1}^r f_k x_k}{\sqrt{\sum_{k=1}^r f_k^2}}\right) = \sum_{n_1+n_2+...+n_r=n} \prod_{k=1}^r \left\{\frac{f_k^{n_k}}{n_k!}H_{n_k}(x_k)\right\} \quad (6)$$

Let us multiply both sides by $H_n(x_{r+1})$ and sum by $n$ using the following equation [11]:

$$\sum_{k=0}^{\infty}\frac{x^k H_k(y)H_k(z)}{2^k k!} = \frac{1}{\sqrt{1-x^2}}\exp\left(\frac{2xyz - (y^2+z^2)x^2}{1-x^2}\right) \quad (7)$$

Let us add normalization and exponential multipliers to transform Hermit polynomials into oscillator base functions. Then we get the following Schmidt decomposition:

$$\left[\sum_{n_1,...n_r=0}^{\infty} c_{n_1,...n_r}\psi_{n_1}^{(1)}(x_1)..\psi_{n_r}^{(r)}(x_r)\psi_{n_1+...+n_r}^{(r+1)}(x_{r+1})\right]^2 = \frac{1}{(\sqrt{\pi})^{r+1}}\exp\left(-\sum_{i,j=1}^{r+1}A_{ij}x_i x_j\right) \quad (8)$$

The right side of the equation is the density of $r+1$-dimensional Gauss distribution.

Matrix $A$ elements are defined by:

$$A_{ij} = \delta_{ij} + \frac{2 f_i f_j}{\left(1-f^2\right)} \qquad i,j = 1,2,...,r,$$

$$A_{i,r+1} = A_{r+1,i} = -\frac{2 f_i}{\left(1-f^2\right)} \quad i=1,2,..,r \quad A_{r+1,r+1} = \frac{1+f^2}{\left(1-f^2\right)} \tag{9}$$

Note that $\det A = 1$.

The validity of the proposed formula for Schmidt decomposition is also verified numerically for particular cases of 2D, 3D, and 4-dimensional Gauss distributions.

Matrix $A$ is positively defined. Out of $r+1$ eigenvalues $r-1$ eigenvalues are equal to unity. The other two are the maximum and the minimum eigenvalues:

$$\lambda_{max} = \frac{\left(1+\sqrt{f^2}\right)^2}{1-f^2} \qquad \lambda_{min} = \frac{\left(1-\sqrt{f^2}\right)^2}{1-f^2} \tag{10}$$

The eigenvalues obviously satisfy the condition:

$$\lambda_{min} \lambda_{max} = 1$$

Let us consider orthonormal eigenvectors of matrix $A$ that describe normal oscillations of the system

To the eigenvalue $\lambda_{max}$ corresponds the following eigenvector

$$a_i^{max} = -\frac{f_i}{\sqrt{2f^2}} \; i=1,2,..,r, \; a_{r+1}^{max} = \frac{1}{\sqrt{2}} \tag{11}$$

Similarly, to the eigenvalue $\lambda_{min}$ corresponds the eigenvector that is different from the latter by the sign of the primary $r$ components.

$$a_i^{min} = \frac{f_i}{\sqrt{2f^2}} \; i=1,2,..,r, \; a_{r+1}^{min} = \frac{1}{\sqrt{2}} \tag{12}$$

The latter $r-1$ eigenvectors that correspond to one eigenvalue equal to unity are constructed as follows. Their $r+1$-th component is equal to zero $a_{r+1} = 0$. The primary $r$ components form the vector that is orthogonal to vector $f$. All these $r-1$ form a basis of a sub-space that is orthogonal complement to vector $f$

Thus, only two out of $r+1$ normal oscillations of the system affect the distinguished $r+1$-th particle. These oscillations correspond to the eigenvlaues $\lambda_{max}$ and $\lambda_{min}$ of matrix $A$. All other oscillations take place inside the $r$-particle system that is formed by the primary $r$ oscillators.

Matrix $A$ can be rewritten as

$$A = VDV^+, \tag{13}$$

where $D$ - is a diagonal matrix of eigenvalues, $V$ - is a unitary matrix which columns are eigenvectors.

Numerical calculations verify the validity of the analytical results presented above for eigenvectors and eigenvalues of matrix $A$.

Matrix of covariance is proportional to matrix $A^{-1}$

$$\Sigma = \frac{1}{2}A^{-1} \tag{14}$$

Let us analyze the probability structure of the weights of states in Schmidt decomposition. Let us conduct a measurement of the distinguished $r+1$-th particle, The considered particle is a measure of the total energy of the other $r$ particles. Let the particle be in state $n$. The probability of that event is:

$$P_n = (1-f^2)(f^2)^n \tag{15}$$

The probabilities $P_n$ satisfy the normalization condition

$$\sum_{n=0}^{\infty} P_n = 1 \tag{16}$$

The probabilities $P_n$ form a geometric progression similar to the state of heat equilibrium of a harmonic oscillator. The corresponding effective temperature can be obtained from the condition

$$\exp\left(-\frac{\hbar\omega}{\theta}\right) = f^2, \tag{17}$$

Therefore, the temperature is:

$$\theta = -\frac{\hbar\omega}{\ln(f^2)} \tag{18}$$

Note that the obtained distribution $P(n) = g^n(1-g)$ (here $g = f^2 = \exp\left(-\frac{\hbar\omega}{\theta}\right)$) is geometric distribution. Its mean value is described by Planck equation.

$$\bar{n} = \bar{n}_1 + \bar{n}_2 + ... + \bar{n}_r = \frac{g}{1-g} = \frac{1}{\exp\left(\frac{\hbar\omega}{\theta}\right) - 1} \tag{19}$$

As a result we get a micro-detector ($r+1$-th particle) in state $n$ and the initial vector transforms to the state where

$$n_1 + n_2 + ... + n_r = n \tag{20}$$

Thus, measurement upon the distinguished $r+1$-th particle defines the total energy of the other $r$ particles. Further measurement of the remaining $r$-particle partial state leads us to multinomial distribution

$$P(|n_1\rangle,|n_2\rangle,...,|n_r\rangle) = p_1^{n_1} p_2^{n_2} ... p_r^{n_r} \frac{(n_1 + n_2 + ... + n_r)!}{n_1! n_2! ... n_r!}, \tag{21}$$

where $p_1 = \frac{f_1^2}{f^2}, p_2 = \frac{f_2^2}{f^2}, ..., p_r = \frac{f_r^2}{f^2}$

It is evident that $p_1 + p_2 + ... + p_r = 1$

The distribution obtained defined the probability that the first particle is in state $|n_1\rangle$, the second particle in state $|n_2\rangle$ and so on (if the total excitation energy of the $r$-particle system is equal to $\hbar\omega(n_1 + n_2 + ... + n_r) = \hbar\omega n$ )

Let us consider all possible outcomes of measurements when $n$ is not fixed but rather described by heat distribution (15). Then using generating functions we get a multi-dimensional distribution of occupation numbers in the system of $r$ oscillators that exactly corresponds to the heat equilibrium conditions. Joint distribution of $r$ particle states is defined by

$$P(k_1, k_2, ..., k_r) = (1-g)\frac{(k_1 + k_2 + ... + k_r)!}{k_1! k_2! \cdots k_r!} g^{(k_1+k_2+...+k_r)} p_1^{k_1} p_2^{k_2} ... p_r^{k_r}, \quad (22)$$

where $g = \exp\left(-\frac{\hbar\omega}{\theta}\right)$

For an arbitrary $j$-th oscillator the probability distribution $P_j(k)$ for different energy states is defined by a geometric distribution

$$P_j(k) = \pi_j (1 - \pi_j)^k \quad (23)$$

where $\pi_j = \frac{1-g}{1 - gq_j}$, $q_j = 1 - p_j$, $k = 0, 1, ...$

Mean value of the considered random variable is

$$M(k) = \bar{n}_j = \frac{p_j}{\exp\left(\frac{\hbar\omega}{\theta}\right) - 1} \quad (24)$$

By analogy mean energy of excitation of an oscillator is:

$$\bar{\varepsilon}_j = \hbar\omega\bar{n}_j = \frac{\hbar\omega p_j}{\exp\left(\frac{\hbar\omega}{\theta}\right) - 1} \quad (25)$$

The latter two equations correspond to the Planck equation. Here we set some weight $p_j$ to every oscillator (in the most simple case all weight can be equal).

**2. Physical model**

Let us consider a physical model that can produce the entangled EPR states described above. Let $r$ oscillators be attached to the partition as shown on Fig.1. Due to the finite mass of the partition the oscillators become coupled with each other. Also an additional $r+1$-th oscillator can be attached from the external side that may serve as a measure device for the system.

First of all we solve a classical problem of normal oscillations of the considered system. It will be shown that there are only two normal oscillations that involve $r+1$-th particle. Schmidt base sets described above will correspond to the oscillations. With vacuum state measurement in Schmidt basis automatically statistical physics distributions arise (in particular, Planck distribution).

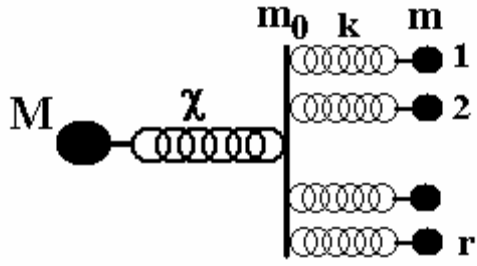

$k, m$ - rigidity and mass of oscillators with numbers $j = 1, 2, ..., r$

$\chi, M$ - rigidity and mass of $r+1$-th oscillator

$m_0$ - partition mass

Fig 1. A system of oscillators, coupled by a partition.

For investigating classical oscillations of the system let us construct its Lagrange function. Let $x_0$ - be the partition coordinate, $z_1, z_2, ..., z_r, z_{r+1}$ - deformations of the corresponding springs. Then coordinates of oscillators are

$$x_1 = x_0 + z_1, \; x_2 = x_0 + z_2, ..., \; x_r = x_0 + z_r, \; x_{r+1} = x_0 - z_{r+1} \quad (26)$$

The minus sign in the latter equation is due to the fact that for the $r+1$-th particle extension of the springs is equal to its coordinate reduce.

The Lagrange function of the system is

$$L = \frac{m_0 \dot{x}_0^2}{2} + \sum_{i=1}^{r} \frac{m(\dot{x}_0 + \dot{z}_i)^2}{2} + \frac{M(\dot{x}_0 - \dot{z}_{r+1})^2}{2} - \sum_{i=1}^{r} \frac{k z_i^2}{2} - \frac{\chi z_{r+1}^2}{2} \quad (27)$$

The Lagrange equation for $\dot{x}_0$ (partition movement) leads to inertial motion of the system as a whole (law of momentum conservation)

$$m \sum_{i=1}^{r} \dot{z}_i - M \dot{z}_{r+1} + (M + rm + m_0)\dot{x}_0 = P_0 = const \quad (28)$$

The other $r+1$ Lagrange equations are:

$$m \ddot{z}_i - \frac{m^2}{M + rm + m_0} \sum_{j=1}^{r} \ddot{z}_j + \frac{mM}{M + rm + m_0} \ddot{z}_{r+1} = -k z_i, \quad (29)$$
$i = 1, 2, ..., r$

$$\frac{M(rm + m_0)}{M + rm + m_0} \ddot{z}_{r+1} + \frac{Mm}{M + rm + m_0} \sum_{j=1}^{r} \ddot{z}_j = -\chi z_{r+1} \quad (30)$$

Let us introduce a dimensionless parameter proportional to the square of frequency

$$\lambda = \frac{\omega^2 m}{k}$$

Normal oscillations of the system we will describe by the eigenvalues of parameter

$$\lambda_j = \frac{\omega_j^2 m}{k} \quad j = 1, 2, ..., r, r+1 \quad (31)$$

Two values of parameter $\lambda$ that describe coupled oscillations of the system and the measurement device are solutions of the following equation

$$a\lambda^2 + b\lambda + c = 0,$$ (32)

where

$$a = \frac{(r\,m + m_0)}{m} - \frac{M\,r}{(M + m_0)},$$ (33)

$$b = -(M + r\,m + m_0)\left(\frac{\chi}{k}\frac{1}{M} + \frac{(r\,m + m_0)}{m(M + m_0)}\right),$$ (34)

$$c = \frac{\chi}{k}\frac{(M + r\,m + m_0)^2}{M(M + m_0)}$$ (35).

The roots are:

$$\lambda_{1,2} = \frac{-b \pm \sqrt{b^2 - 4ac}}{2a}$$ (36)

The two considered normal oscillations have a very simple form – all of the primary $r$ particles have the same displacement $z_1 = z_2 = ... = z_r$, while the $r+1$ - th particle has the displacement

$$z_{r+1}^{(1,2)} = \frac{(M + r\,m + m_0)/\lambda_{1,2} - (M + m_0)}{M} z_1^{(1,2)}$$ (37)

Other $r-1$ values of parameter $\lambda$ are equal to unity.

$$\lambda_j = 1, \quad j = 3, 4, ..., r+1$$

From the results it is evident that for an infinite large mass partition ($m_0 \to \infty$), take place independent oscillations of oscillators with masses $m$ and $M$ correspondingly. In that case:

$\lambda_1 = \frac{\omega_1^2 m}{k} = 1$, therefore $\omega_1^2 = \frac{k}{m}$ - are oscillations of an oscillator of mass $m$;

$\lambda_2 = \frac{\omega_2^2 m}{k} = \frac{\chi\,m}{kM}$, therefore $\omega_2^2 = \frac{\chi}{M}$ - are oscillations of an oscillator of mass $M$

In other boundary case, when the coupling of the measurement device with the system is small ($\chi \to 0$) we have $\lambda_1 = \frac{\omega_1^2 m}{k} = 1 + \frac{rm}{m_0}$, $\lambda_2 = \frac{\omega_2^2 m}{k} = 0$ (38)

As we see the frequencies of the system in this case do not depend on the mass of the measurement device $M$. In this case $r$ - particle system and the partition form an entangled state. During the process of measurement of the state statistical distribution arises that corresponds to the heat equilibrium. The corresponding temperature is given by the equation:

$$\theta = \frac{\hbar\omega}{\ln\left(\frac{(1+\xi)^{3/2} + 1}{(1+\xi)^{3/2} - 1}\right)}, \text{ where } \xi = \frac{rm}{m_0}$$ (39)

The considered dependence is illustrated on Fig.2 where the temperature and excitation energy are presented as a function of the ratio of the partition mass to the system mass.

The field where the considered graphs are close can be considered as classical. It corresponds to small values of partition mass and corresponds to high temperatures.

For smaller temperatures the considered oscillator degrees of freedom are freezing out and the graph of excitation energy is below the temperature graph. Note that the larger the mass of the partition the more quantum become the observed probability distributions. For the infinite large mass of the partition the particles do not entangle with it which corresponds to zero temperature.

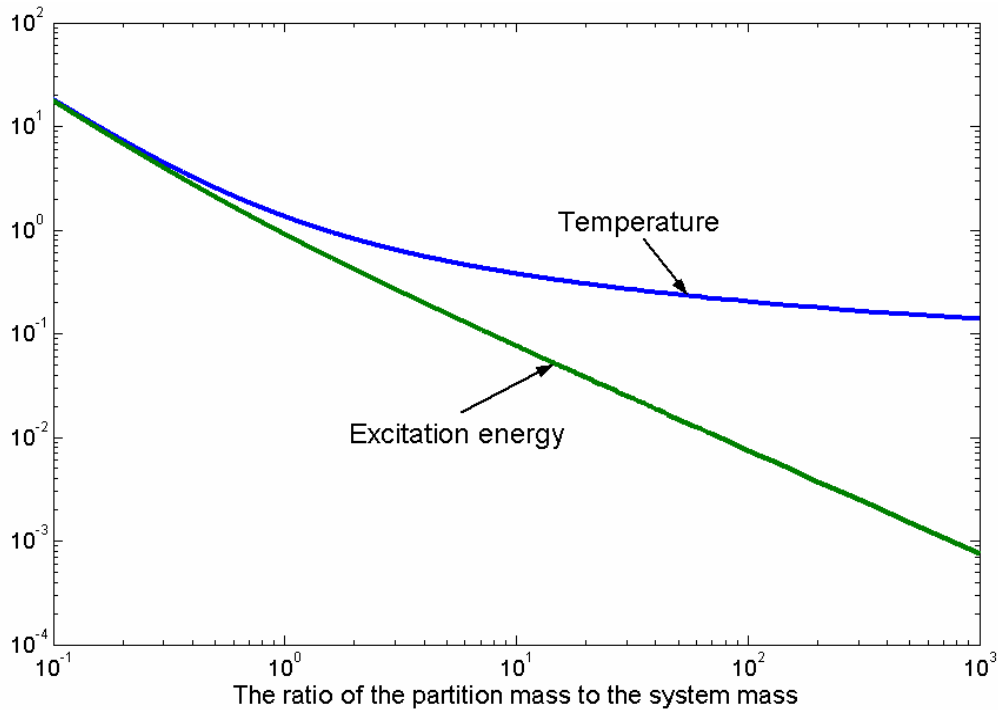

Fig.2. Ann illustration of the dependence of the temperature and excitation energy on the ratio of the partition mass to the system mass.

The importance of the obtained result is that the temperature of the system is an internal property of the system that can be explicitly calculated even for systems with a large number of degrees of freedom. (Before, the value has been phenomenological).

The characteristic temperature of the system $\theta_0$ that is similar to the temperature of Einstein and Debye in solid state physics can be derived for the condition $\xi = \dfrac{rm}{m_0} \approx 1$ (where all of the oscillators taken together serve as an partition). Then $\theta_0 = 1.3532\,\hbar\omega$. Note that in general the considered model leads to the results that are similar to the Einstein's theory of specific heats. The development of the model will be the matter of further research.

In principle the relation between entanglement and temperature can be expressed as follows. Consider a multi-particle system at zero temperature (all normal oscillators are in the ground state). Then during a transition from normal oscillators to initial physical oscillators, the process of measurement upon the oscillators yields in the value of the temperature close to the characteristic temperature of the system. Therefore, the measurement (interaction with environment) results in the "revival" of the system. If the system is then left alone, it will again moves to the ground state (zero temperature). On the contrary, if the system is open and it interacts with the environment then its temperature is defined by the intensity of the interaction.

For a more intense interaction the temperatures are higher and the system is more classical; for weaker interactions the system is closer to the ground state (low temperatures).

Parameters of the measurement device (mass and rigidity of the $r+1$-th oscillator) can be chosen so that the corresponding vacuum state of $r+1$ particles is an EPR-state of type (1). Then for the measurement the following condition takes place $n_1 + n_2 + ... + n_r = n$.

Let us introduce operators (matrices) of mass and rigidity. The elements of the mass matrix are

$$\hat{\mu}_{ii} = m\left(1 - \frac{m}{M + rm + m_0}\right), \quad i = 1,...,r$$

$$\hat{\mu}_{ij} = -\frac{m^2}{M + rm + m_0}, \quad i \neq j, \quad i,j = 1,...,r$$

$$\hat{\mu}_{i,r+1} = \hat{\mu}_{r+1,i} = \frac{Mm}{M + rm + m_0}, \quad i = 1,...,r$$

$$\hat{\mu}_{r+1,r+1} = \frac{M(rm + m_0)}{(M + rm + m_0)} \tag{40}$$

The rigidity matrix is diagonal

$$\hat{\kappa}_{ii} = k \quad i = 1,...,r$$

$$\hat{\kappa}_{r+1,r+1} = \chi$$

$$\hat{\kappa}_{ij} = 0 \quad i \neq j \quad i,j = 1,...,r+1 \tag{41}$$

Both matrices are symmetrical and positively defined.

Eigenvalues of the system and corresponding oscillations are the roots of the following problem [12]:

$$\omega_s^2 \hat{\mu} z_s = \hat{\kappa} z_s \quad s = 1,...,r+1 \tag{42}$$

Here $\omega_s$ and $z_s$ - are frequency and amplitude vector of the $s$ - th normal oscillation

To solve the problem one has to find such matrices $V_1$ and $D_1$, so that:

$$\hat{\kappa} V_1 = \hat{\mu} V_1 D_1. \tag{43}$$

Here $D_1$ - is a diagonal matrix of eigenvalues, while columns of $V_1$ are eigenvectors

Therefore matrix $A$ that corresponds to the vacuum state (8) is equal to

$$A = V_1 \sqrt{D_1} V_1^+ \tag{44}$$

Normal oscillations are mutually orthogonal. Therefore, [12]:

$$z_s^+ \hat{\mu} z_j = 0, \quad z_s^+ \hat{\kappa} z_j = 0 \quad s \neq j, \quad s,j = 1,...,r+1 \tag{45}$$

Normal oscillations allow one to transform a system of $r+1$ coupled oscillators to a system of $r+1$ independent oscillators. The mass and rigidity of every normal oscillator is defined as:

$$\mu_s = z_s^+ \hat{\mu} z_s, \quad \kappa_s = z_s^+ \hat{\kappa} z_s, \quad s = 1,...,r+1 \tag{46}$$

Numerical solution of the problem verifies the validity of the obtained results.

**Conclusion:**

Let us formulate the main results of the work:

1. Multi-particle entangled EPR-type states based on oscillator functions are constructed. One of the oscillator particles is distinguished and it may be considered as a measure of the total energy of other particles.
2. It is demonstrated that thermodynamic equilibrium may be considered as an effect of quantum entangling of the system vacuum state. It is worth noting that we do not introduce any statistical mechanism different from that in the postulates of quantum mechanics to describe thermodynamic distributions.
3. An explicit physical multi-particle model is developed and studied. In the model entanglement takes place in the process of interaction of oscillator particles by a mutual intermediary. The proposed model allows one to discover the nature and origins of equilibrium quantum distributions. The temperature of the system is defined by the ratio of the particles' mass to the intermediary mass. Thus, the introduced notion of temperature is not phenomenological, but can rather be calculated as a physical parameter of the system.
4. The results of provided analytical researches are in close accordance with the results of numerical calculations.

The paper has been supported by the Russian Foundation for Basic Research grant (06-07-89129)